\documentclass[aip,
 amsmath,amssymb,
 reprint,%
]{revtex4-1}

\usepackage{graphicx}
\usepackage{dcolumn}
\usepackage{bm}

\usepackage[utf8]{inputenc}
\usepackage[T1]{fontenc}
\usepackage{mathptmx}
\usepackage{etoolbox}

\usepackage{bm}
\usepackage[colorlinks=true,linkcolor=blue]{hyperref}
\expandafter\ifx\csname package@font\endcsname\relax\else
 \expandafter\expandafter
 \expandafter\usepackage
 \expandafter\expandafter
 \expandafter{\csname package@font\endcsname}
\fi
\def\bq{\begin{equation}}
\def\eq{\end{equation}}
\def\bqy{\begin{eqnarray}}
\def\eqy{\end{eqnarray}}

\def\p{\partial}

\def\p{\partial}

\makeatletter
\def\@email#1#2{%
 \endgroup
 \patchcmd{\titleblock@produce}
  {\frontmatter@RRAPformat}
  {\frontmatter@RRAPformat{\produce@RRAP{*#1\href{mailto:#2}{#2}}}\frontmatter@RRAPformat}
  {}{}
}%
\makeatother
\begin{document}

\preprint{AIP/123-QED}

\title[Hall MHD waves: A fundamental departure from their MHD counterparts]{Hall MHD waves: A fundamental departure from their MHD counterparts}
\author{Swadesh M. Mahajan}
\email{mahajan@mail.utexas.edu}
\affiliation{Department of Physics and Institute for Fusion Studies, The University of Texas at Austin, Austin, TX 78712, USA}
\affiliation{Department of Physics, School of Natural Sciences, Shiv Nadar University, Uttar Pradesh 201314, India}

\author{Prerana Sharma}
\email{prerana.sharma@austin.utexas.edu}
\affiliation{Physics Department, Ujjain Engineering College, Ujjain, 
Madhya Pradesh 456010, India}

\author{Manasvi Lingam}
\email{mlingam@fit.edu}
\affiliation{Department of Aerospace, Physics and Space Sciences, Florida Institute of Technology, Melbourne, FL 32901, USA}
\affiliation{Department of Physics and Institute for Fusion Studies, The University of Texas at Austin, Austin, TX 78712, USA}

\date{}%

\begin{abstract}
It is demonstrated through a succinct derivation as to how the linear waves in Hall magnetohydrodynamics (HMHD) constitute a fundamental departure from the standard MHD waves. Apart from modifying the conventional MHD spectrum, the Hall current induces a distinct and new branch consisting of purely circularly polarized waves that may become the representative shear waves. 
\end{abstract}

\maketitle


It is thoroughly established that ideal (or resistive) magnetohydrodynamics (MHD) is a cornerstone of plasma physics \cite{KT73,TS97,BS03,JPF14,SG16,GB17,GKP19}, with applications ranging from space science and astrophysics to nuclear fusion and aerospace engineering. It is important, however, to recall that MHD is not a universally valid theory, as it is derived under a particular set of assumptions \cite{JPF14}, which apply to a subset of plasmas.

In particular, when the plasma is chiefly collisionless and we focus on phenomena occurring at length scales comparable to, or smaller than, the ion skin depth, it is necessary to go ``beyond MHD'' and avail ourselves of the models comprising extended MHD \cite{GKP19}. One of the most crucial models in this regard is Hall MHD (HMHD), which entails the inclusion of a Hall current term within the MHD induction equation. HMHD has been utilized in a variety of contexts to uncover diverse phenomena, as briefly summarized later.

The basics of HMHD were delineated and elaborated decades ago \cite{Light60,RTC79,EW86,Tur86,Holm87}, and a bevy of equilibrium solutions have been subsequently derived \cite{MY98,YM99,MY00,YMOIS01,TT06,BKS09,GKT24}. Variational and Hamiltonian formulations of MHD have attracted extensive attention \cite{Yosh02,HYM,BKS12,YH13,KLMWW14,ML15,AKY15,DAML16,LMM15,LMM16}, partly because they have enabled formal analysis of the underlying mathematical structure of HMHD, such as its conserved quantities. In tandem, a multitude of papers on waves in HMHD have analyzed its mathematical characteristics \cite{SBG03,OM04,HIM,IHM04,HII05,Gal06,SGB07,FV09,JZZ14,AY16}; of this category, a special class of waves, termed linear-nonlinear waves was the subject of Refs. \cite{MK05,MM09}. 

From an astrophysical standpoint (e.g., protoplanetary disks), HMHD has been invoked in conjunction with the famous magnetorotational instability (MRI) to analyze how the former influences the evolution of the latter \cite{BT01,Ward07,Bai14,LB16}. Other applications of HMHD to real-world systems include generic energy conversion mechanisms \cite{MMNS01,OSYM01,KM10}; large- and small-scale dynamo processes \cite{MGM02,MAP07,GMD10,Ling16,HBC23}; the unified dynamo-reverse dynamo mechanism \cite{MSMS05,LingMa15}; small-scale turbulence \cite{GSR96,KM04,GB07,BCX15,FGS19,PFL19,BSC20,ML20,PY22,BH24}; fast reconnection \cite{AB04,Fitz04,DSS08,ZY09,YKJ10,BKS11,CB16,MY22}; and even newly emerging disciplines like astrobiology \cite{LL19,ML21}. We mention, however, that fast magnetic reconnection may also be mediated by resistive MHD mechanisms such as the plasmoid instability \cite{TS97,BHY09,CLHB,CLH17}. Many of these studies were conducted under the incompressible assumption, which is physically reasonable in select circumstances and can help simplify the analytical or numerical modeling.

Hence, it is apparent that HMHD is a key model in plasma physics. Thus, elucidating its fundamental properties would be of much significance. This short note accordingly demonstrates, through a readily accessible calculation, that linear waves associated with incompressible HMHD are considerably more interesting and involved than what seems to be generally realized and appreciated. Given that HMHD remains a very active area of investigation both for linear and nonlinear studies (as indicated in the preceding paragraphs), a compact -- yet essentially complete -- delineation of the fundamental waves (normal modes) of the incompressible HMHD system ought to be valuable to myriad researchers. 

In a nutshell, this endeavor of working out the salient physics of (incompressible) HMHD waves is the chief objective of our current work -- even if some/all aspects of the derivation were to have been reported before, at the minimum, this pedagogical treatment serves to synthesize together disparate ``branches'' of incompressible HMHD waves into a coherent whole, and may additionally offer novel insights into the nature of these waves and their deviations from their ideal MHD counterparts.

We plunge straight into the calculation. In Alfv\'enic units (the magnetic field normalized to some ambient value $B_0$, velocities to the Alfv\'en speed $V_A= B_0/\sqrt{ 4\pi m_{i} {n}}$, lengths to some system length $L$, and the frequencies normalized to $V_A/L$), the
equations of incompressible HMHD are expressible as 
\begin{equation} \label{Binit}
{
\frac{\p {\bf B}}{\p t} =   \nabla \times \left[\left({\bf V} - \epsilon {\bf J}\right) \times {\bf B}\right]},
\end{equation}
\begin{equation} \label{Vinit}
{
 \frac{\p {\bf V}}{\p t} = {\bf V} \times \left(\nabla \times {\bf V}\right) + {\bf J} \times {\bf B} - \nabla \left(\frac{{\bf V}^2}{2}\right)},
\end{equation}
where ${\bf J} \equiv \nabla \times {\bf B}$, and $\epsilon \equiv \lambda_{i}/L$ is the normalized skin depth (since $\lambda_i$ is the ion skin depth) and measures the magnitude of the Hall current contribution to the system. We proceed with the assumption of zero pressure, whose validity is not strictly preserved. The inclusion of a finite pressure would amount to the inclusion of an extra term inside the gradient term (i.e., last term on the RHS) in Eq. (\ref{Vinit}). To put it differently, the net effect is that certain waves would acquire a contribution to the dispersion relation arising from $C_s \neq 0$, where $C_s$ denotes the sound speed, as touched on later.

Let us split the dynamical variables into the equilibrium component(s) plus fluctuations,
$${{\bf B} = \hat{e}_z + {\bf b },\quad  {\bf V} = {\bf v}} $$
with the ambient field of magnitude unity (in normalized units) aligned along the $z$ direction; this choice can be made without loss of generality, because we are free to select the coordinate system. For this very simple conceptual problem, there is no ambient flow, as is the case in many traditional studies of plasma waves; in other words, we have a static equilibrium. On taking the linear limit (where the fluctuations are small), the fluctuating fields then evolve as 
\begin{equation}\label{Bwave}
{
\frac{\p {\bf b}}{\p t}  + \epsilon\left(\hat{e}_z \cdot \nabla\right) {\bf j} - \nabla \times \left({\bf v} \times \hat{e}_z\right)=0 }
\end{equation} 
\begin{equation} \label{Vwave}
{
 \frac{\p {\bf v}}{\p t} - {\bf j} \times \hat{e}_z=0 }
\end{equation}
where ${\bf j} \equiv \nabla \times {\bf b}$. Taking the time derivative of Eq. (\ref{Bwave}), substituting Eq. (\ref{Vwave}), and after simplification we end up with
\begin{equation} \label{Nlinwav}
{
 \frac{\p^2 {\bf b}}{\p t^2} - \nabla^2 {\bf b}  - \nabla j_z \times \hat{e}_z = - \epsilon \frac{\p}{\p t} \left(\hat{e}_z \cdot \nabla {\bf j}\right)}
\end{equation}
We remark that the linear system -- namely Eq. (\ref{Nlinwav}) -- is expressed fully in terms of magnetic quantities; it can, as a matter of fact, conveniently be expressed solely in terms of $b_z = \hat{e}_z \cdot {\bf b}$, and $j_z = \hat{e}_z \cdot {\bf j}$. {After taking the scalar product of Eq. (\ref{Nlinwav}) with $\hat{e}_z$, and using appropriate identities from vector calculus (e.g., ${\bf X} \cdot \left({\bf X} \times {\bf Y}\right) = 0$), we end up with}
\begin{equation} \label{CompMod}
{\left[\frac{\p^2}{\p t^2} - \nabla^2\right] b_z = - \epsilon\frac{\p}{\p t} \left(\hat{e}_z \cdot \nabla\right) j_z },
\end{equation}
{and on applying the operator $\hat{e}_z \cdot \left(\nabla \times \right)$ to Eq. (\ref{Nlinwav}), and simplifying the resultant equation via vector calculus, we have}
 \begin{equation} \label{ShearMod}
 {\left[\frac{\p^2}{\p t^2} - \left(\hat{e}_z\cdot\nabla\right)^2\right] j_z  =  \epsilon \frac{\p}{\p t} \left(\hat{e}_z \cdot \nabla\right) \nabla^2 b_z}.
\end{equation}
that might appear (we will see later it is not quite so) to fully describe HMHD waves; it reduces to conventional MHD for $\epsilon=0$. The latter (in the linear approximation carried out so far) allows two totally independent modes (we introduce $k_z \equiv {\bf k} \cdot \hat{e}_z$, which is the wave vector along the equilibrium field):
\begin{enumerate}
    \item The compressional mode ($D \equiv \omega^2 - k^2=0$) corresponding to the existence of finite $b_z$ and $j_z=0$.  
    \item The shear Alfv\'en wave ($F \equiv \omega^2 - k_z^2=0$), with $b_z=0$ and finite current ($j_z\neq0$).
\end{enumerate}
These two modes are completely along expected lines, thus serving as a consistency check.

Next, by turning the Hall current on, and taking the Fourier transform, Eqs. (\ref{CompMod}) and (\ref{ShearMod}) reduce to
 \begin{equation}\label{Disp1}
  {D b_z = i \epsilon k_z \omega ({\bf k}\times{\bf b})\cdot {\hat{e}_z}}
\end{equation}
\begin{equation}\label{Disp2}
 { F ({\bf k}\times{\bf b})\cdot {\hat{e}_z} = -i \epsilon k_z {\omega} k^2 b_z}
\end{equation}
where $j_z=i\hat{e}_z\cdot \left({\bf k}\times{\bf b}\right)$ has been used, which follows from the fact that ${\bf j} = i\left({\bf k} \times {\bf b}\right)$ in the Fourier space. When we probe into deeper aspects (for instance, polarization) of HMHD waves, we will need to employ the Fourier transforms of Eqs. (\ref{Bwave}) and (\ref{Vwave}), which are respectively given by
\begin{equation} \label{RelA}
 {\omega {\bf b} = - k_z {\bf v} + \hat{e}_z {\bf k}\cdot{\bf v} + i \epsilon k_z \left({\bf k} \times {\bf b}\right)}, 
 \end{equation}
 \begin{equation}\label{RelB}
 {\omega {\bf v} = - k_z {\bf b} + b_z {\bf k}}.
\end{equation}
Taking the scalar product of Eq. (\ref{RelB}) with ${\bf k}$, and using the relation ${\bf k} \cdot {\bf b} = 0$ in Fourier space, leads us to
 \begin{equation}\label{RelComp}
 {b_z =\frac{\omega {\bf k}\cdot{\bf v}}{k^2}},  
\end{equation}
and likewise, substituting Eq. (\ref{RelB}) in Eq. (\ref{RelA}) yields
\begin{equation} \label{RelA2}
{F {\bf b} - b_z ( k^2 \hat{e}_z-k_z {\bf k} )= i \epsilon \omega k_z \left({\bf k} \times {\bf b}\right)}, 
 \end{equation}

Next, by starting from Eqs. (\ref{Disp1}) and (\ref{Disp2}), we take their product to arrive at the well-known dispersion relation:
\begin{equation}\label{Disp}
 { FD= (\omega^2 - k_z^2)( \omega^2 - k^2)=\epsilon^{2} {k_z}^{2} {\omega}^{2} k^2}
\end{equation}
where the shear and the compressional branch are linearly coupled by the Hall current (to wit, equivalent to specifying nonzero $\epsilon$). This dispersion relation is the zero plasma pressure limit (i.e., tantamount to setting $C_s = 0$ for the sound speed), for instance, presented and/or elaborated in Refs. \cite{OM04,HIM,IHM04} (also refer to Refs. \cite{SBG03,SGB07}). If we had included the pressure, a finite $C_s$ would have entered the equation accordingly.

One of the key distinguishing features of (incompressible) HMHD waves is evident from the relations Eqs. (\ref{Disp1}) and (\ref{Disp2}). Unlike in the case of linear MHD waves, the quantities $b_z$ and $({\bf k}\times{\bf b})\cdot {\hat{e}_z} \propto j_z$ are no longer independent and must either be zero or nonzero together. In other words, the Hall current lashes together the temporal evolution of the $z$-components of the magnetic field and of the current density; in consequence, HMHD, which represents a singular perturbation of ideal MHD \cite{Holm87,OY05,Gal06}, opens the door for novel physics. This striking facet of the coupling is attested from a careful consideration of either Eqs. (\ref{RelComp}) and (\ref{RelA2}), or Eqs. (\ref{Disp1}) and (\ref{Disp2}). 

When both these quantities ($b_z$ and $j_z$) are nonzero, we may duly proceed with the dispersion relation given by Eq. (\ref{Disp}). However, a more radical feature induced by the the Hall current is the existence of an entirely new class of modes for which $b_z = 0 = ({\bf k}\times{\bf b})\cdot {\hat{e}_z}$ -- it is this class of modes that we will chiefly dwell on in this paper. Note that this solution ($b_z = 0 = ({\bf k}\times{\bf b})\cdot {\hat{e}_z}$) is fully consistent with Eqs. (\ref{Disp1}) and (\ref{Disp2}), because it simultaneously converts the RHS and LHS of these equations to zero. We can approach $b_z = 0$ through an alternative path: the incompressibility of the velocity field corresponds to $\nabla \cdot {\bf v} = 0$, which yields ${\bf k} \cdot {\bf v} = 0$ in the Fourier domain. On plugging this expression in Eq. (\ref{RelComp}), it is verifiable that $b_z = 0$.

Therefore, when the condition 
\begin{equation}\label{current1}
 { j_z= ({\bf k}\times{\bf b})\cdot {\hat{e}_z}= k_x b_y- k_ y b_x = 0 }
\end{equation}
is coupled with the divergence condition of $\nabla \cdot {\bf b} = {\bf k} \cdot {\bf b} = 0$ (after setting $b_z=0$):
\begin{equation}\label{div1}
 {  k_ x b_x +   k_y b_y = 0 },
\end{equation}
we are forced to demand either that 
 \begin{equation}\label{Belt1}
 {  b^{2}_ x   +   b^{2}_y  = 0 },
\end{equation}
is valid, or instead that we have
\begin{equation}\label{oneD1}
{  k^{2}_ x   +   k^{2}_y  = 0 }
\end{equation}
The condition inherent in Eq. (\ref{oneD1}) either limits the system to one dimension  ($k_x=0=k_y$) or imparts an imaginary component to the wave vector, and will not be discussed; we point out that the latter scenario would translate to an exponentially growing or damping mode (in space).

We will thereupon concentrate on Eq. (\ref{Belt1}), which proves to be exactly the condition exhibited by circularly polarized waves, as intimated shortly. In addition, since we are working with $b_z=0$, combining this with Eq. (\ref{Belt1}) yields
\begin{equation}\label{BeltHMHD}
     { b_x^2 + b_y^2 + b_z^2 = 0 \quad \Rightarrow \quad {\bf b}^2 = 0,}
\end{equation}
{where the second expression follows from recognizing that ${\bf b} \cdot {\bf b} \equiv {\bf b}^2 = b_x^2 + b_y^2 + b_z^2$ when written in terms of $b_x$, $b_y$, and $b_z$. However, at the same time, it is important to appreciate at this stage that ${\bf b}$ is implicitly a complex-valued vector. Thus, we may also express it explicitly in the canonical form:}
\begin{equation}\label{bBreakdown}
    {{\bf b} = {\bf b}_R \pm i {\bf b}_I,}
\end{equation}
{where ${\bf b}_R$ and ${\bf b}_I$ are the real and imaginary components of this complex-valued vector, respectively (but both of which are individually real-valued vectors). On calculating ${\bf b}^2$ from the aforementioned equation (i.e., treating it as the square of a complex vector), and setting it to zero as per the second relation in Eq. (\ref{BeltHMHD}), we end up with the two results:}
\begin{equation}\label{CompBfield}
    {|{\bf b}_R|= |{\bf b}_I|, \quad  {\bf b}_R \cdot {\bf b}_I =0.}
\end{equation}
{Another avenue for directly arriving at Eq. (\ref{CompBfield}) is to use $b_z = 0$ from earlier along with Eq. (\ref{Belt1}), which jointly implies that}
\begin{equation}
     {{\bf b} = b_x \hat{e}_x + b_y \hat{e}_y = b_x \hat{e}_x \pm i b_x \hat{e}_y,}
\end{equation}
{where the second equality is derived from invoking $b_y = \pm i b_x$, which is itself a consequence of Eq. (\ref{Belt1}). Thus, on comparing the aforementioned equation with Eq. (\ref{bBreakdown}), we see that ${\bf b}_R = b_x \hat{e}_x$ and ${\bf b}_I = b_x \hat{e}_y$, which automatically leads us to Eq. (\ref{CompBfield}). In general, as revealed from an inspection of Eq. (\ref{CompBfield}), ${\bf b}_R$ and ${\bf b}_I$ are two mutually perpendicular vectors of equal magnitude.}

Next, recalling that $b_z = 0$ and substituting this expression in Eq. (\ref{RelA2}), we duly obtain
\begin{equation} \label{RelA2Rem}
{F  {\bf b} = i \epsilon \omega k_z \left({\bf k} \times {\bf b}\right)}, 
 \end{equation}
which is tantamount to the Fourier transform of a Beltrami equation with circularly polarized waves (see discussion later) obeying the dispersion relation:
\begin{equation} \label{RelA2Disp}
{F^{2} = \epsilon^{2} {k_z}^{2} {\omega}^{2} k^2,} 
\end{equation}
To see why the statement below Eq. (\ref{RelA2Rem}) is warranted, we emphasize that a Beltrami equation is given by $\nabla \times {\bf X} = \mathcal{C} {\bf X}$ \cite{MY98,YM99,YMOIS01,ML15}, where $\mathcal{C}$ is conventionally a constant and ${\bf X}$ is the vector field of interest; a vector field satisfying this Beltrami equation is known as a Beltrami field. Hence, the Fourier transform yields ${\bf k} \times {\bf X} = -i \mathcal{C} {\bf X}$, which is apparently identical in structure to Eq. (\ref{RelA2Rem}).

{It is worth noting that Eq. (\ref{RelA2Disp}) has been obtained by taking the scalar product of Eq. (\ref{RelA2Rem}) with the complex conjugate of the same Eq. (\ref{RelA2Rem}), and then using the fact that ${\bf k} \cdot {\bf b} = 0$.} Next, on extracting the square root of Eq. (\ref{RelA2Disp}), we have
\begin{equation} \label{RelA2Remv2}
{F = \pm \epsilon k_z \omega k
 }, 
\end{equation}
that can be readily solved to yield
\begin{equation} \label{DispBelt}
{ \frac{\omega}{k_z} = \pm \frac{\epsilon  k}{2} \pm \left(1+ \frac {\epsilon^{2} k^2}{4}\right)^{\frac{1}{2}} \approx \pm1\pm \frac{\epsilon  k}{2},
}
\end{equation}
where the last equality is valid for the limit of $\epsilon k \ll 1$.

With the derivation of the linear wave dispersion relations of incompressible HMHD out of the way, we are in a position to provide a deeper perspective, and underscore the inherent significance of these findings:
\begin{itemize}
\item Perhaps the most important of all is the emergence of a totally new branch of purely circularly polarized (Beltrami) waves/modes exhibiting the polarization of
\begin{equation} \label{polBelt}
{{\bf b}= {\bf b}_R \pm i {\bf b}_I,  \quad  |{\bf b}_R|= |{\bf b}_I|, \quad  {\bf b}_R \cdot {\bf b}_I =0}, 
\end{equation}
\begin{equation} \label{vbz}
{{\bf v}= -\frac{k_z}{\omega} {\bf b},  \quad   b_z=0},
\end{equation}
which are rendered possible only via the Hall current; {\it this branch does not exist in ideal MHD}. {It is the expression of the fact that the Hall term is a singular perturbation that imparts novel physics to HMHD with respect to ideal MHD. The formulation of this singular Beltrami branch has been outlined earlier: note that Eq. (\ref{polBelt}) is an outcome of the text between Eqs. (\ref{BeltHMHD}) and (\ref{RelA2Rem}),} whereas Eq. (\ref{vbz}) is obtained from Eq. (\ref{RelB}) after setting $b_z = 0$ in this expression. {These modes are Beltrami in nature, since they satisfy the Beltrami equation as highlighted in Eq. (\ref{RelA2Rem}) and subsequent discussion.}

\item The circularly polarized waves referenced earlier are endowed with unique mathematical characteristics -- conspicuously different from those of the standard linear waves -- because, for each individual wave (viz., for a given ${\bf k}$), the nonlinear terms in the HMHD wave system can be demonstrated to vanish exactly \cite{MK05}. Thus, even though our analysis entailed a linear derivation, these waves are actually wave solutions of HMHD with arbitrary amplitude, and have been discussed in several papers under the nomenclature of linear-nonlinear waves \cite{MK05,MM09,AY16,ALM16}. In this regard, these modes ought to be perceived as the HMHD counterparts of the Alfv\'en-Walen solutions ${\bf b} = \pm {\bf v}$, which are arbitrary amplitude wave solutions documented in ideal MHD \cite{CW44,HA50}.

\item The aforementioned linear-nonlinear waves exhibit independence from the waves represented by the dispersion relation Eq. (\ref{Disp}) despite the seeming partial resemblance of Eqs. (\ref{Disp}) and (\ref{RelA2Disp}). Let us, for instance, examine Eq. (\ref{Disp}) for what would be approximately a shear Alfv\'en wave (${\omega}^{2} \approx k^{2}_z$) in MHD. For the sake of simplicity, we shall compare the two sets of waves in the regime of $k^2 \gg k^{2}_z$ and $\epsilon k \ll 1$. The first set of Eqs. (\ref{Disp1}), (\ref{Disp2}), and (\ref{Disp}) collectively yields
\begin{equation} \label{DispSimple}
{ \frac{\omega}{k_z} = \pm \frac{1}{\sqrt{1+\epsilon^{2} {k_z}^{2}}}
}
 \end{equation}
with the polarization (after some simple algebra):
\begin{equation} \label{RelA2Remv3}
{ b_x= -\frac{k_x}{k^{2}_x+k^{2}_y} \left[k_z+\frac{i k^{2}}{\epsilon \omega k_z}\right]b_z},
\end{equation}
\begin{equation} \label{RelA2Remv4}
{ b_y= -\frac{k_y }{k^{2}_x+k^{2}_y} \left[k_z-\frac{i k^{2}}{\epsilon \omega k_z}\right]b_z
},
\end{equation}
which is manifestly distinct from the nonlinear-linear waves, to wit, the set of Beltrami modes described by Eqs. (\ref{DispBelt}), (\ref{polBelt}), and (\ref{vbz}).
 \end{itemize}
 
These two branches of linear modes are, loosely speaking, mutually orthogonal to each other, implying that no linear transitions can take place among them; for instance, the Beltrami modes associated with the linear-nonlinear waves exhibit $b_z = 0$, whereas the other class of modes evinces $b_z \neq 0$, as is evident from inspecting Eqs. (\ref{RelA2Remv3}) and (\ref{RelA2Remv4}). 

The two sets of waves each have their own pros. The linear-nonlinear wave solutions have been used to explain the potential energy spectra in solar wind turbulence through Kolmogorov-type scaling arguments \cite{ALM16}, as well as to explain an observed absence of equipartition (of magnetic and kinetic energy) in Hall turbulence and dynamos \cite{LB16,Ling16}. Likewise, the other set of waves have been invoked in understanding certain properties of Hall turbulence \cite{SBG03,OM04,HIM,IHM04,HII05,SGB07}.

This phenomenon would be essentially ported over to the more general case of extended MHD when electron inertia, for example, is included; the reason is that the Hall current still remains and induces analogous effects, albeit with (smaller) corrections stemming from the finite electron mass and skin depth \cite{MM09,AY16,ALM16}. Hence, all studies ``beyond MHD'' (i.e., with two-fluid contributions) must properly account for the fundamentally altered structure of the linear modes, in fact, of the incompressible branch that constitutes a powerful new addition to the repertoire of Alfv\'enic waves.

\section*{Acknowledgments}
This work was supported by U.S. DOE under Grants Nos. DE- FG02-04ER54742 and DE-AC02-09CH11466.

\nocite{*}
\bibliography{HallMHDwaves}

\end{document}